\documentclass[pss]{wiley2sp} % provides pss two-column style
\usepackage{amsmath}
\usepackage{float}
\usepackage{afterpage}
\usepackage{placeins}
\newcommand{\cdlq}{c_{\text{L}q}^{\dagger}}
\newcommand{\clq}{c_{\text{L}q}}

\newcommand{\dd}{d^{\dagger}}

\newcommand{\elq}{\varepsilon_{\text{L}q}}
\newcommand{\epo}{\varepsilon_0}
\newcommand{\G}{\Gamma}
\newcommand{\la}{\langle}
\newcommand{\ma}[1]{{\mathcal #1}}
\newcommand{\ra}{\rangle}
\newcommand{\tl}{{\rm L}}
\newcommand{\tr}{{\rm R}}
\newcommand{\vlq}{V_{\text{L}q}}

\begin{document}

% Title of the article
\title{Power spectrum of  electronic heat current fluctuations}

% Abbreviated title for the page headers
\titlerunning{Short title }

% Authors
\author{%
  Fei Zhan\textsuperscript{\textsf{\bfseries 1,2}},
  Sergey Denisov\textsuperscript{\textsf{\bfseries 1}},
  and Peter H\"anggi\textsuperscript{\Ast\textsf{\bfseries 1,3}}}

% Abbreviated list of authors for the page headers
\authorrunning{First author et al.}

%E-mail-address of corresponding author
\mail{e-mail:
  \textsf{hanggi@physik.uni-augsburg.de}, Phone:
  +49-821-5983249, Fax: +49-821-5983222}

% author's affiliations/addresses
\institute{%
  \textsuperscript{1}\,Institut f\"ur Physik, Universit\"at Augsburg,
Universit\"atsstr.~1, D-86159 Augsburg, Germany\\
  \textsuperscript{2}\,Centre for Engineered Quantum Systems, School of Mathematics and Physics, The
University of Queensland, St Lucia QLD 4072, Australia\\
  \textsuperscript{3}\,Department of Physics and Centre for Computational
Science and Engineering, National University of Singapore, Singapore 117546, Singapore\\
  }

\received{XXXX, revised XXXX, accepted XXXX} % do not change, will be filled in by the publisher
\published{XXXX} % do not change, will be filled in by the publisher

% Please select about four verbal keywords for your manuscript.

\abstract{%
% This is a macro for the typesetting of two-column text in an
% abstract. It will typeset the two arguments in \abstcol{}{} as the
% left and right column inside the abstract box. At the
% columnbreak there will be always a columnbreak (\par), so both
% columns start with a new paragraph. No automatic column height
% balancing is done.
%
% If used with a \titlefigure it will silently output both
% parameters as consecutive paragraphs.
%
% The macro is defined exclusively inside the argument of \abstract{};
% if used outside it will raise an error.
%
% Usage: \abstcol{<left column>}{<right column>}
\abstcol{%
 We analyze the fluctuations of an electronic thermal current across an idealized molecular junction.
The focus here will be on the spectral features
of the resulting  heat fluctuations. By use of the Green function method we derive an explicit
expression for the frequency-dependent power spectral density of the emerging energy fluctuations. The complex expression
simplifies considerably in the limit of zero frequency, yielding the noise intensity of the heat current.
The spectral density for the electronic heat fluctuations still depends on the frequency in the zero-temperature limit,
assuming different asymptotic behaviors in the low- and high-frequency regions. We further address  subtleties
and open problems from an experimental viewpoint for  measurements of frequency-dependent power spectral densities}.
}

% The class file requires the standard graphicx Latex package. See the 'LaTeX
% standard graphics and color packages documentation' for more information at
% <http://tug.ctan.org/tex-archive/macros/latex/required/graphics/grfguide.pdf>.
%
% Accepted figure file formats depend on which LaTeX flavour is used.
% Classic LaTeX is always able to use Encapsulated Postscript (EPS);
% PDFLaTeX can't use this but accepts PDF, JPG, PNG, and GIF formats.
%
% See examples for implementing graphics in floating figure environments later in this file.
% If \titlefigure is given, it takes as its mandatory parameter the
% % name (without extension) of some figure file.
%\begin{figure}[b]
%\includegraphics[width=5cm]{icon}
\titlefigure[]{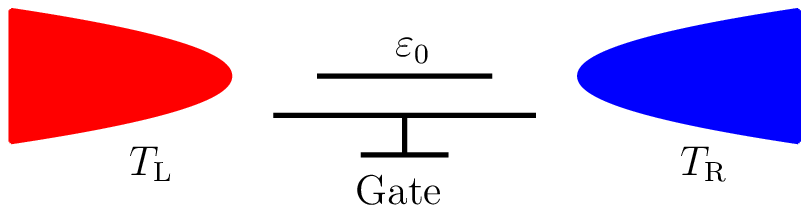}
%\end{figure}
\titlefigurecaption{\label{pssb.201349192_Fig_1.eps}(Color online)
Sketch of a molecular junction setup used in the text. The average heat flow is generated by electrons moving from a hot electrode $T_{\rm L}$
across the molecular junction towards a neighboring cold electrode $T_{\rm R}$.
The inter-electrode electronic level $\varepsilon_{0}$ can be tuned continuously.\vfill%\vspace{1truecm}}
}
\maketitle   % please do not remove

\section{Introduction}

 The experimental activities over the last fifteen years in investigating electronic transport across molecular junctions~\cite{reed97science,nitzan2003science,joachim2005pnas}
have triggered  several waves of intense research in  theory
\cite{blanter00,hanggi2002,kohler2005pr,cuniberti,dubi11rmp} and experiment \cite{joachim00nature,taonj2006nt,cuixd2001s,reichert02prl}.
Single molecule electronics is still considered as a promising candidate for the substitution of  silicon-based elements in
the information processing technology~\cite{nitzan2003science,joachim2005pnas,joachim00nature,taonj2006nt}. Likewise, molecular junctions have advantages
in the context of energy-related applications. This is due to the potential of hybrid solid-state molecular structures which enable
novel interface features. Moreover, the abundant selection of possible molecules and electrode materials  allow to tailor specific
properties. In particular,  the topic of thermoelectric \cite{dubi11rmp} and photovoltaic \cite{photo} conversion processes
continue to prompt timely research  in the field of molecular electronics.

Apart from the standard  current-voltage characteristics~\cite{reed97science,cuixd2001s,reichert02prl}, it is also
important to obtain insight into the fluctuations that accompany the corresponding transport processes. For example, by
use of the full counting statistics~\cite{levitov04prb,morten08prb,bagrets03prb,esposito} it is possible
to extract information about the fluctuations of the electric current flowing across a
molecular wire \cite{clement07prb,beenakker03,liyuanp90apl,buttiker92prb,camalet04prb}.

In the context of thermoelectric applications, the issue of energy  transport through molecular
junctions and the properties of the corresponding fluctuations acquire special importance. Thermal fluctuations may crucially
impact the electronic transport features, and even affect the overall performance of the molecular junction.
With the molecular systems  operating on the nanoscale  corresponding energy current fluctuations can become sizable.
This may be so even in situations where the average energy current is {\it vanishing} identically, as it is the case in thermal
equilibrium with both interconnecting electrodes held at the same temperature. Moreover, the properties of nonequilibrium noise correlations,
or likewise, its frequency-dependent spectral properties,  are in no obvious manner related to the mean value of the energy flow itself.
With this work we shall  explore the fluctuations of the heat current caused by the transferring electrons.
Our goal is to obtain analytical estimates for the {\it power spectral density} (PSD) of the  heat fluctuations, even at the
expense that these may mainly apply to idealized setups only. With such a restriction these analytical results may nevertheless be useful
to appraise the role of heat current noise in more realistic molecular junctions. It is further of interest to have an estimate available
when devising molecular circuitry  for more complex tasks.

Energy transport across a molecular structure which links two electrodes is induced by a difference of the two electrode temperatures, see in Fig. \ref{fig:model}. The physics of heat transfer generally involves both
electrons and phonons and their mutual interaction
\cite{dubi11rmp,zhanfei2009pre,sivan86prb,koch04prb,chenyuchang2005prl,galperin06prb,galperin07prb,segal06prb,paulsson03prb,galperin07jpcm,wangjiansheng08epjb}.
Therefore, the amount of energy flow carried across the wire should be addressed with care, with the
need to distinguish between energy transfer mediated either by electrons or phonons, or a combination of both.
If phonons are mainly at work this situation relates to the new field of {\it phononics} \cite{linianbei2012rmp}, a novel research area
which may lead to new circuit elements, such as
molecular thermal diodes, thermal transistors, thermal logic gates, to name but a few
~\cite{linianbei2012rmp,lehmann1,lehmann2,changcw06science,wanglei07prl,libaowen04prl,libaowen06apl}.
Then, the size of fluctuations in heat current does matter; this is so because those may well turn out to be deleterious to
intended information processing tasks.

Heat transport mediated by electrons  relates at the same time to charge transfer:
electrons  moving from lead-to-lead  carry not only charge but also energy ~\cite{ea81prb,sivan86prb,b90jpcm}.
However, the amount of energy transferred by a single electron, unlike to its charge, is not quantized~\cite{rey07prb}.
In contrast to those studies that examine the average heat flow, however, much less attention
has been paid to the issue of \textit{fluctuations} of the accompanying flow of energy.
In  prior work~\cite{krive01prb} the energy transport through a ballistic quantum wire has been considered in the Luttinger-liquid limit,
by neglecting the discreteness of the wire's energy spectrum. Likewise, with Refs.~\cite{averin10prl}, the PSD of the
heat current fluctuations has been derived within a scattering theory approach, using the assumption that the electrons are transmitted (reflected) at
the same rate, independently of their actual energies. The results of the last two papers, however, are
challenging because it has been shown therein that the noise characteristics of heat current at equilibrium exhibits a well-pronounced frequency
dependence even at absolute  zero-temperature.  Therefore, this very zero temperature finding is in contradiction with the naive  expectation as  provided by the equilibrium fluctuation-dissipation theorem (FDT). This found deviation from the FDT in those works is  attributed loosely to the role of zero-point-energy fluctuations \cite{averin10prl}.

With this work we shall consider the {\it electronic} energy current that proceeds across a molecular wire composed of a single energy level with the two electrodes held at  different temperatures. A preliminary short discussion of such electronic nonequilibrium heat noise  has been presented by us with Ref.~\cite{Feiprb2011}. Here, we complement and extend this study and present further useful details on the theoretical derivation of the noise expression.  Moreover, we discuss the  nonequilibrium heat noise of the corresponding heat current over much broader parameter regimes and frequency regimes {\it away} from the zero-frequency limit. With our setup we also corroborate  the results obtained in the zero temperature limit for the power spectral density at finite frequencies for a different setup in Ref.  \cite{averin10prl}.  In addition, we address several subtleties when it comes to the explicit validation of our theoretical findings by experimental means.

\section{Molecular junction setup\\}

In order to obtain analytical tractable expressions we shall neglect electron-phonon interactions and, as well, electron-electron interactions. Such a simplification can be justified for tailored situations that involve a very short wire only. Then, the Coulomb interaction via a double occupancy shifts the energy far above the Fermi level so that its role in thermal transport can be neglected. Likewise, the electron dwell time is short as
compared to the electron-phonon relaxation time scale.
Note however, that in contrast to previous works \cite{averin10prl}, we account here for the dependence of the transmission coefficient on its electron energies, and, within the Green function approach \cite{kohler2005pr,wangjiansheng08epjb}, derive an explicit expression for the PSD of the heat current fluctuations, $\tilde{S}^{\rm h}(\omega)$. In particular we demonstrate below that the net noise features of the heat current are quite distinct from their electronic counterpart.

%With this result at hand we explore different regimes of electron transport and demonstrate that the
%heat noise in fact is quite distinct from its electric counterpart.

\begin{figure}[b]
\centering
 \includegraphics[width=.618\linewidth]{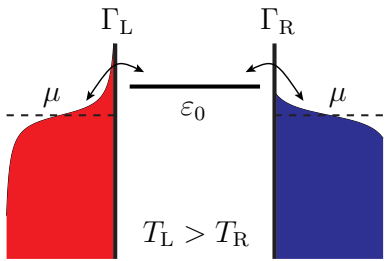}
\caption{\label{fig:model}(color online)~Idealized setup of a molecular junction used in text: Two metal leads, each filled with an ideal electron gas, are connected by a single orbital $\epo$. The
coupling strengths are determined by $\G_{\text{L/R}}$. The left lead is prepared at a higher temperature as compared to the opposite right
lead, i.e. $T_{\text{L}} > T_{\text{R}}$. The chemical potential, $\mu$, is the same for both leads so that no electric current due to a finite voltage bias is present.}
\end{figure}

Our  molecular junction setup is depicted with Fig.~\ref{fig:model}: It is described by a Hamiltonian
\begin{equation}
 H=H_{\text{wire}}+H_{\text{leads}}+H_{\text{contacts}}\;. \label{totalham}
\end{equation}
It contains three different contributions, namely the wire Hamiltonian, the leads and the wire-lead coupling, respectively. We consider here the regime
of {\it coherent} quantum transport whereby neglecting dissipation inside the wire. The wire is composed of a single orbital; i.e.,
\begin{equation}
H_{\rm wire}=\epo\dd d\;,\label{hammol}
\end{equation}
at an energy $\epo$, with the fermionic creation and annihilation operators, $\dd$ and $d$. The energy level $\epo$ can be tuned by applying a gate voltage. Our idealized setup allows for explicit analytical calculations. Physically, it mimics a double barrier resonant tunneling structure
$\text{GaAs/Al}_x\text{Ga}_{1-x}$-structure of the type considered for electronic shot noise calculations in Ref. \cite{bo96jpcm}, herein truncated to a single
Landau level. As commonly implemented, the electrodes are modeled by reservoirs, composed of ideal electron gases, i.e.,
\begin{equation}
H_{\text{leads}}=\sum_{\ell q}\varepsilon_{\ell q}c_{\ell q}^{\dagger}c_{\ell q}\;,\label{hamld}
\end{equation}
where the operator $c_{\ell q}^{\dagger}(c_{\ell q})$ creates (annihilates) an electron with momentum $q$ in the $\ell=$L (left) or $\ell=$R (right) lead. We
assume that the electron distributions in the leads are described by the grand canonical ensembles at the temperatures $T_{\rm L/R}$ and with chemical
potentials $\mu_{\rm L/R}$. Using such ideal electron reservoirs we obtain
\begin{equation}
\label{ensembleav}
\la c_{\ell q}^{\dagger}c_{\ell'
q'}\ra=\delta_{\ell\ell'}\delta_{qq'}f_{\ell}(\varepsilon_{\ell q}) \;,
\end{equation}
where
\begin{equation}
f_{\ell}(\varepsilon_{\ell q})=\left[e^{(\varepsilon_{\ell q}-\mu_{\ell})/k_{\text{B}}T_{\ell}}+1\right]^{-1}
\end{equation}
denotes the Fermi function.

We impose a finite temperature difference $\Delta T=T_{\text{L}}-T_{\text{R}}$ and use identical chemical potentials, $\mu_{\rm L}=\mu_{\rm R} = \mu$ for the electrodes. When an electron tunnels out from a lead, the energy $E$ is transferred into the wire which presents the heat transfer, $\delta Q$. Observing the value for the chemical potential, $\mu$, it reads $\delta Q = (E-\mu)$. In the following we use that all the electron energies are measured from the chemical potential value $\mu$, being set at $\mu=0$.

The Hamiltonian which describes the tunneling events reads:
\begin{equation}
 H_{\text{contacts}}=\sum_{\ell q}V_{\ell q}c_{\ell q}^{\dagger}d + h.c.\;.\label{hamcon}
\end{equation}
This part mediates the coupling between the wire and the electrodes. Here, the notation $h.c.$ denotes Hermitian conjugate. The quantity
$V_{\ell q}$ is the tunneling matrix element, and the tunneling coupling is characterized in general by a spectral density,
\begin{equation}
\Gamma_{\ell}(E)=2\pi\sum_{q}|V_{\ell q}|^2\delta(E-\varepsilon_{\ell q}) \;. \label{specden}
\end{equation}

In the following, we shall use a wide-band limit of the electrode conduction bands, setting $\Gamma_{\ell}(E):=\Gamma_{\ell}$.

\section{Power spectral density of electronic heat current fluctuations\\}

 Working within the Heisenberg description of operators we present the detailed derivation of the electronic energy
current induced by a finite temperature difference of the two leads and the PSD of the corresponding energy fluctuations.
We  limit the consideration to pure energy transfer that proceeds in absence of a finite voltage
bias across the two leads and no  particle concentration across the leads. Put differently, no cross-phenomena of energy transfer due
to a charge current (i.e. no Joule heating) or  due to a particle concentration  current (i.e. no Dufour effect) is at work.
Therefore, because  all other channels for the energy transport between the leads  are then explicitly excluded  form our  consideration,
we follow
previous works, e.g. see in Refs.~\cite{ea81prb,averin10prl},  and use throughout this study the term `heat current' as synonym for energy current.
The electronic thermal current  then reads
\begin{equation}
J^{\rm h}_{\rm L} (t)=\frac{\sum\delta Q(t)}{\Delta t} \;.
\end{equation}

With our choice of chemical potentials $\mu_L=\mu_R=0$, we find that the heat transfer operator is $\delta Q (t)=E_{\rm L}$, with the energy operator given by
\begin{equation}
E_{\tl}=\sum_{q}\elq\cdlq\clq\;.
\end{equation}
Its time derivative thus yields the operator for the heat flux, reading:
\begin{equation}
J^{\rm h}_{\tl}(t)=-\sum_q\frac{2\elq}{\hbar}\text{Im}[\vlq\cdlq(t) d(t)]\label{jlop}\; .
\end{equation}
The heat current is positive valued when heat transport proceeds from the hot left lead, i.e. $T_{\rm L} > T_{\rm R}$ to the adjacent cold lead, see in Fig. \ref{fig:model}. In deriving the above expression we have employed the Heisenberg
representation for the lead electron operators. The average current is obtained by the ensemble average
$\la J^{\rm h}_{\rm L}(t)\ra$.  Because there are no electron sinks and sources in between the leads
we have $\la J^{\rm h}_{\tl}(t)\ra=- \la J^{\rm
 h}_{\tr}(t) \ra$. We henceforth focus on the quantities derived with regard to
the left lead; i.e.,  $\la J^{\rm h}_{\rm L}(t)\ra := \la J^{\rm h}(t)\ra$.

The quantum correlation function of heat current fluctuations is described by the {\it symmetrized} auto-correlation function, i.e.,
\begin{equation}
\label{fluxcorrelation}
S^{\rm h}(t,t')=\frac{1}{2}\left\la[\Delta J^{\rm h}_{\tl}(t),\Delta J^{\rm h}_{\tl}(t')]_+\right\ra\;,
\end{equation}
with respect to the operator of the heat current fluctuation
\begin{equation}
\Delta J^{\rm h}_{\tl}(t)=J^{\rm h}_{\tl}(t)-\la J^{\rm
  h}_{\tl}(t)\ra\; .
\end{equation}

The heat current noise is described  with $\tau = t-t'$ by the {\it symmetrized} quantum auto-correlation function
\begin{equation}\label{sdef}
S^{\rm h}(\tau)=1/2\la[\Delta J^{\rm h}_{\ell}(\tau),\Delta J^{\rm h}_{\ell}(0)]_+\ra\;,
\end{equation}
of the heat current fluctuation operator $\Delta J^{\rm h}_{\ell}(s)=J^{\rm h}_{\ell}(s)-\la J^{\rm h}_{\ell}(s)\ra$, where the anti-commutator $[A,B]_+=AB+BA$ ensures the Hermitian property.

With this work we throughout consider the asymptotic long time limit $t\rightarrow\infty$ when all transients are decayed.
In this asymptotic limit the
average heat current is stationary and the auto-correlation function of the heat current fluctuations becomes time-homogeneous; i.e. it is independent of initial preparation effects. It thus depends on the time difference $\tau=t-t'$ only. The Fourier transform yields the power spectral density (PSD) $\tilde{S}^{\rm h}(\omega)$ for the heat current noise, i.e.,
\begin{equation}\label{ss}
\tilde{S}^{\rm h}(\omega) = \tilde{S}^{\rm h}(-\omega) = \int_{-\infty}^{\infty} d\tau e^{i\omega\tau}S^{\rm h}(\tau) \geq 0\;.
\end{equation}
$\tilde{S}^{\rm h}(\omega)$ is an even function in frequency and
strictly semi-positive, in accordance with the Wiener-Khintchine theorem \cite{HanggiPR82}.
In the following we address positive values of the frequency, $\omega > 0$, only.\\

The annihilation operators of the electrode
states satisfy the Heisenberg equations of motion; i.e.,
\begin{equation}
\dot{c}_{\ell q}(t)=-\frac{i}{\hbar}\varepsilon_{\ell q}c_{\ell
  q}(t)-\frac{i}{\hbar}V_{\ell q}d(t) \; ,
\label{heisenberg}
\end{equation}
yielding the solution
\begin{align}
c_{\ell q}(t)=&c_{\ell q}(t_0)e^{-i\varepsilon_{\ell q}(t-t_0)/\hbar}\notag\\
&-\frac{iV_{\ell q}}{\hbar}\int_{t_0}^{t}dt'e^{-i\varepsilon_{\ell q}(t-t')/\hbar}d(t') \;.
\label{clq}
\end{align}
Here, the first term on the right hand side describes the dynamics of the free electrons in the leads, while the second term accounts for the influence of the molecule.

The Heisenberg equation of the molecular annihilation operator is given by
\begin{equation}
\dot{d}(t)=-\frac{i}{\hbar}\varepsilon_0d(t)-\frac{i}{\hbar}\sum_{\ell q}V^*_{\ell q}c_{\ell q}(t).\label{moloper}
\end{equation}
Upon inserting Eq. \eqref{clq} into Eq.~\eqref{moloper}, we obtain
\begin{equation}
\dot{d}=\frac{i}{\hbar}\epo d(t)-\frac{\G_{\rm L}+\G_{\rm R}}{2\hbar}d(t)+\xi_{\rm L}(t)+\xi_{\rm R}(t),\label{cem}
\end{equation}
where we have defined the noise operator
\begin{equation}
\xi_{\ell}(t)=-\frac{i}{\hbar}\sum_qV^*_{\ell q}\exp\left[-\frac{i}{\hbar}\varepsilon_{\ell q}(t-t_0)\right]c_{\ell q}(t_0).\label{opergausnoise}
\end{equation}
In addition, we have employed the definition \eqref{specden} and used the wide-band limit.

The noise quantity defined in Eq.~\eqref{opergausnoise} denotes
operator-valued Gaussian noise, which is characterized by its mean
 and correlation properties, reading
\begin{align}
\la\xi_{\ell}(t)\ra &= 0\\
\la\xi^{\dagger}_{\ell'}(t')\xi_{\ell}(t)\ra &=
\delta_{\ell\ell'}\int_{-\infty}^{\infty}\frac{d\varepsilon}{2\pi\hbar^2}e^{-i\varepsilon(t-t')/\hbar}\Gamma_{\ell}(\varepsilon)f_{\ell}(\varepsilon).
\end{align}

This noise accounts for the influence of the states stemming from the electrodes $l = {\rm L},{\rm R}$.

Now the central problem is to solve the inhomogeneous differential equation~\eqref{moloper}. Once we obtain the solution of Eq.~\eqref{moloper}, we obtain also the solution for Eq.~\eqref{clq}, the heat current \eqref{jlop} and also the power spectral density in Eq.~\eqref{ss}.

To obtain the solution of Eq.~\eqref{cem}, we follow the Green function approach in Ref.~\cite{kohler2005pr} and start with solving the following differential equation
\begin{equation}
 (\frac{d}{dt}+\frac{i\varepsilon_0}{\hbar}+\frac{\G_{\rm L}+\G_{\rm R}}{2\hbar})G(t-t')=\delta(t-t')\;, \label{green}
\end{equation}
followed by the application of the convolution $d(t)=\int G(t-t')(\xi_{\tl}(t')+\xi_{\tr}(t'))dt'$.
The solution of Eq. \eqref{green} is thus given by:
\begin{equation}
 G(t)=\theta(t)e^{-i\epo t/\hbar-(\G_{\rm L}+\G_{\rm R})t/2\hbar}\;.
\end{equation}
Then, the molecular operator in Eq.~(\ref{cem}) assumes the form
\begin{equation}\label{ct}
d(t)=\sum_{\ell q} V^*_{\ell q}\frac{\exp[-i\varepsilon_{\ell q}(t-t_0)/\hbar]}{\varepsilon_{\ell q}-\epo+i(\G_{\rm L}+\G_{\rm R})/2}c_{\ell q}(t_0)\;.
\end{equation}\\

\begin{figure}[t]
\centerline{\includegraphics[width=6cm]{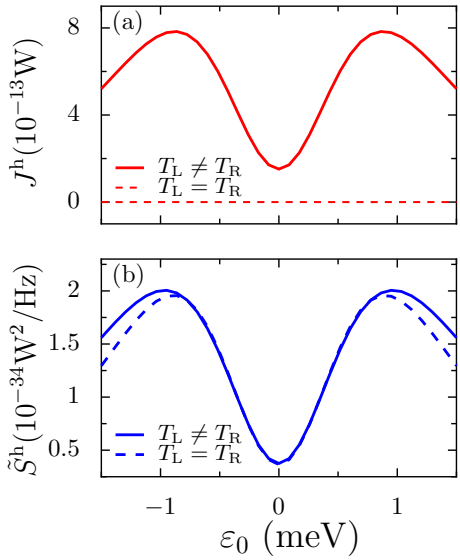}}
\caption{\label{fig:curnoise}  (color online) (a) Average electronic heat current $J^{\rm h}$ and (b)
the zero-frequency values of corresponding heat noise power $\tilde{S}^{\rm h}$ of the accompanying heat current fluctuations
as a function of orbital energy for an identical lead coupling strength $\Gamma=0.1$~meV. The  parameters are:
$T_\text{L}=5.2$~K, $T_\text{R}=3.2$~K (solid lines) and $T_{\rm L} = T_{\rm R} = 4.2$~K (dashed line).
Figure in parts adapted from Ref. \cite{Feiprb2011}.}
\end{figure}

In what follows we address solely the asymptotic properties which are reached with the initial time of preparation $t_0\rightarrow - \infty$.
This implies that average currents assume stationary values and correlation functions become time-homogeneous.
With this expression and its Hermitian conjugate, we obtain the occupation value of the molecular energy level $\epo$ as
\begin{align}\label{epo}
\overline{n}_{\epo}=&\,\la d^{\dagger}(t)d(t)\ra\notag\\
=&\sum_{\ell\ell'qq'}\frac{V_{\ell q}\exp[i\varepsilon_{\ell q}(t-t_0)/\hbar]}{[\varepsilon_{\ell q}-\varepsilon_0-i(\Gamma_{\tl}+\Gamma_{\tr})/2]} \notag\\
&\times\frac{V_{\ell'q'}^*\exp[-i\varepsilon_{\ell'q'}(t-t_0)/\hbar]}{[\varepsilon_{\ell'q'}-\varepsilon_0+i(\Gamma_{\tl}+\Gamma_{\tr})/2]}\la c^{\dagger}_{\ell q}(t_0)c_{\ell'q'}(t_0)\ra\notag\\
=&\sum_{\ell q}\frac{|V_{\ell q}|^2f_{\ell}(\varepsilon_{\ell q})}{(\varepsilon_{\ell q}-\varepsilon_0)^2+(\Gamma_{\tl}+\Gamma_{\tr})^2/4}\;,
\end{align}
where we have employed the ensemble average, Eq.~\eqref{ensembleav}.
We find that this occupation is determined  by the Fermi function of the
leads, weighted by the tunneling matrix elements $V_{\ell q}$ and the difference
between lead states and the  molecular energy level $\varepsilon_0$, see in Eq. (\ref{epo}).
This occupation value is time-independent because there are no time-dependent external fields present.\\

Upon substituting the result in Eq. \eqref{ct} into Eq. \eqref{clq}, we find for the operators in the electrodes
\begin{align}
&c_{\ell q}(t)=c_{\ell q}(t_0)e^{-i\varepsilon_{\ell q}(t-t_0)/\hbar}\notag\\
&+\sum_{\ell' q'}\frac{V_{\ell q}V^*_{\ell' q'}e^{-i\varepsilon_{\ell' q'}(t-t_0)/\hbar}}{\varepsilon_{\ell' q'}-\epo+i(\G_{\rm L}+\G_{\rm R})/2}c_{\ell'
q'}(t_0)\notag\\
&\times B[\varepsilon_{\ell'q'}-\varepsilon_{\ell q}]\;,\label{clqt}
\end{align}
where,
\begin{equation}
B(E)={\mathcal P}\left(\frac{1}{E}\right)-i\pi\delta(E)\;,
\end{equation}
and ${\mathcal P}$ denotes the integral principal value.
In going from Eq.~\eqref{ct} to Eq.~\eqref{clqt} we have used Sokhotsky's formula which states that $\lim_{\epsilon\rightarrow 0}1/(x+i\epsilon)={\mathcal P}(1/x)-i\pi\delta(x)$, where ${\mathcal P}(1/x)=\int_{-\infty}^{0^-}dx/x+\int^{\infty}_{0^+}dx/x$, see in Ref. \cite{Vladimirov}.\\

Next we insert Eq.~\eqref{ct} and Eq.~\eqref{clqt} into the heat current operator, Eq.~\eqref{jlop}, and by
consequently taking the ensemble average, we obtain a Landauer-like formula for the heat
current, reading ~\cite{dubi11rmp,ea81prb,sivan86prb,b90jpcm,galperin07jpcm,segal03jcp}:
\begin{equation}\label{current}
 \la J^{\rm h}(t)\ra := J^{\rm h} = \frac{1}{2\pi\hbar}\int dE E{\mathcal T}(E)[f_{\tl}(E)-f_{\tr}(E)]\;,
\end{equation}
where the transmission coefficient
\begin{equation}
\ma{T}(E)=\G_{\tl}\G_{\tr}/[(E-\epo)^2+\G^2] \;,
\label{transmission}
\end{equation}
is energy-dependent.

The expression for the thermoelectric  charge current~\cite{sivan86prb} reads very similar to Eq.~(\ref{current}), except for its absence of the energy multiplier $E$
in the integral on the rhs of Eq.~(\ref{current}). This seemingly small difference changes, however, the physics of transport through the wire, because
the multiplier inverts the symmetry of the integral. Namely, the thermolelectric current is an antisymmetric function of orbital energy and vanishes when the orbital
energy level is aligned to the chemical potentials of the leads \cite{Feiprb2011}, while the heat current is a symmetric function and acquires a nonzero value at $\varepsilon_0 = 0$, see in Fig.~\ref{fig:curnoise}(a). \\

\subsection{Main result and discussion\\}
\begin{figure}
\centerline{\includegraphics[width=.418\textwidth]{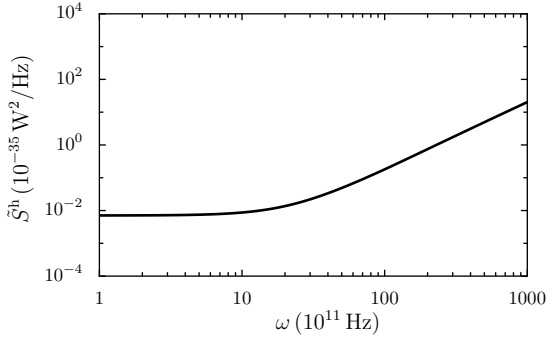}}
\caption{\label{fig:noiseomega}Power spectral density of the heat current noise as a function of the frequency $\omega$ at temperatures $T_{\rm L}=6\,{\rm K}, T_{\rm R}=2\,{\rm K}$. The other parameters are $\varepsilon_0=0$ and $\Gamma=0.1~{\rm meV}$.}
\end{figure}
\begin{figure*}[!h]
\centerline{\includegraphics[width=\textwidth]{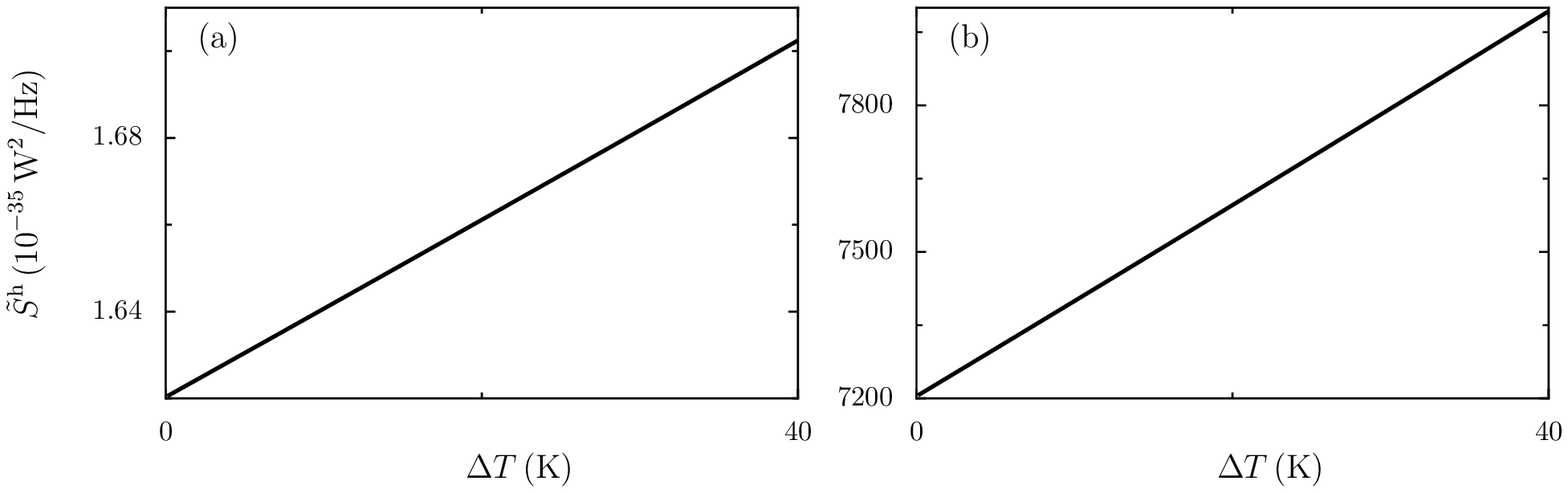}}
\caption{\label{fig:noisedeltatemp}
Power spectral density of heat current noise at frequency
$\omega=2.16\times10^{13}\,{\rm Hz}$ (which is the Debye cut-off
frequency of gold) as a function of temperature difference with (a)
weak molecule-wire coupling $\Gamma=0.1\,{\rm meV}$ or (b) strong
molecule-wire coupling $\Gamma=10\,{\rm meV}$. The other employed
parameters are $T_{\rm R}=300~{\rm K}$ and $\epo=0$.}

\end{figure*}
Upon combining Eq.~\eqref{ss} and Eq.~\eqref{jlop}, we end up after a cumbersome evaluation with the nontrivial expression for the PSD of electronic heat current noise. Due to the complexity of this resulting expression the physics
it inherits is not very illuminative. Nevertheless, we depict it here as given in our preliminary report \cite{Feiprb2011}, reading:
\begin{align}
&\tilde{S}^{\rm h}(\Omega= \hbar\omega; T_{\tl},T_{\tr})\notag\\
=&\sum_{\pm}\int\frac{dE}{4\pi\hbar}\left\{\left[\left(E\pm\frac{\Omega}{2}\right)^2\ma{T}(E)\ma{T}(E\pm\Omega)\right.\right.\notag\\
&+\left.\frac{\G_{\rm L}^2\left[E(E-\epo)-(E\pm\Omega)(E\pm\Omega-\epo)\right]^2}{\left[(E-\epo)^2+\G^2\right][(E\pm\Omega-\epo)^2+\G^2]}\right]\notag\\
&\;\;\times f_{\rm L}(E)\overline{f}_{\rm L}(E\pm\Omega)\notag\\
&+\left(E\pm\frac{\Omega}{2}\right)^2\ma{T}(E)\ma{T}(E\pm\Omega)f_{\rm R}(E)\overline{f}_{\rm R}(E\pm\Omega)\notag\\
&+\left[\left(E\pm\frac{\Omega}{2}\right)\left(\pm\frac{\Omega}{2}\right)\frac{\G_{\rm L}^2\ma{T}(E\pm\Omega)}{(E-\epo)^2+\G^2}\right.\notag\\
&+E^2\ma{R}(E)\ma{T}(E\pm\Omega)\mp\left.\frac{1}{2}E\Omega\ma{T}(E)\ma{T}(E\pm\Omega)\right]\notag\\
&\;\;\times f_{\rm L}(E)\overline{f}_{\rm R}(E\pm\Omega)\notag\\
&+\left[\left(E\pm\Omega\right)\left(\pm\frac{\Omega}{2}\right)\right.\ma{T}(E)\ma{T}(E\pm\Omega)\notag\\
&+\left(E\pm\Omega\right)^2\ma{R}(E\pm\Omega)\ma{T}(E)\notag\\
&+\left(E\pm\frac{\Omega}{2}\right)\left(\mp\frac{\Omega}{2}
\right)\left.\frac{\G_{\rm L}^2\ma{T}(E\pm\Omega)}{(E-\epo)^2+\G^2}\right]\notag\\
&\;\;\times f_{\rm R}(E)\overline{f}_{\rm L}(E\pm\Omega)\;,\label{longeqn}
\end{align}
wherein we abbreviated $\Omega\equiv\hbar\omega$, $\overline{f}\equiv1-f$,
and $\ma{R}(E)\equiv1-\ma{T}(E)$ denoting the reflection coefficient.
Below we consider the case of symmetric coupling between the wire and the leads, $\G_{\rm L}=\G_{\rm R}=\G$.

We emphasize here that this heat PSD is a  manifest nonequilibrium result where with a finite temperature bias the result accounts for `heat'-shot noise and, simultaneously nonequilibrium, Nyquist-like  heat noise. Let us next discuss, via graphical means, some general features of the inherent complexity as depicted with Eq.~(\ref{longeqn}).

In Figure~\ref{fig:noiseomega}, we depict the dependence of the  PSD of
heat current fluctuations versus frequency $\omega$ at finite temperature bias, given by $T_{\rm L}=6\,{\rm K}, T_{\rm R}=2\,{\rm K}$.
We deduce from the figure that this nonequilibrium  PSD exhibits different power laws in different frequency regions and grows with increasing frequency.

Moreover, we find that the spectral density strength $\Gamma$ of the wire-lead coupling can change
the dependence of the  heat fluctuation PSD on the parameters.
In Figure~\ref{fig:noisedeltatemp} we depict the PSD as a function of the temperature difference $\Delta T$ over
a wide regime of $\Delta T = 40$~K, both
in the case of weak and strong wire-lead couplings. With weak coupling,  the PSD
is smaller by one order of magnitude and only weakly (i.e. with a small slope) increases with $\Delta T$, see Fig.~\ref{fig:noisedeltatemp}(a).
In contrast, the PSD increases very fast with $\Delta T$ when the coupling is very strong, see in Fig.~\ref{fig:noisedeltatemp} (b).
According to Eq.~\eqref{transmission}, the transmission coefficient becomes wider when $\Gamma$ is larger, such that more electrons,
whose energies deviate stronger from the chemical potential, are allowed to transport across the molecular junction. Therefore,
the PSD becomes strongly enhanced and depends sensitively on $\Delta T$.

It is striking that both dependencies are near perfectly linear over the wide temperature regime of $\Delta T$.
Given the complex structure of the nonlinear nonequilibrium PSD detailed with the lengthy expression in ~{\eqref{longeqn}}
such extended linearity can hardly be expected a priori. The mechanism behind this distinctive feature
is not evident and thus constitutes an interesting issue for further studies.
\\

\subsection{Issues relating to experimental validation \\}
It should also be mentioned here that the explicit verification of quantum mechanical power spectral densities is experimentally not at all straightforward. In clear contrast to the classical case, the symmetrized quantum correlation for heat in Eq.~(\ref{sdef}) presents
no manifest {\it quantum observable} that can be  measured directly, but rather it is a functional operator expression
involving the  time-evolution of the dynamics. This is so because the heat flux operators at different times
do  not commute. In fact,  a quantum mechanical evaluated  PSD can be measured only indirectly via a single-time measurement of a
tailored linear response function, via a corresponding, generally {\it nonequilibrium} quantum fluctuation-dissipation relation
which connects this response function with a corresponding quantum mechanical two-time correlation expression \cite{HanggiPR82}.
Put differently, this tailored response function is then required to relate {\it precisely} to our so calculated nonequilibrium
quantum correlation of heat  fluctuations in ~\eqref{sdef}.
This is so because a direct two-time quantum measurement of two observables  at different times $t$ would then
impact (i.e. it will generally alter) the a priori theoretically determined quantum two-time correlation expression in \eqref{ss};
for further details and similar  pitfalls see also in Refs. \cite{Campisiprl,Campisipre}, where the problem of measuring quantum work poses the same challenge.

The situation becomes more promising when we focus on the {\it zero-frequency}  result of the PSD for heat noise: The  variance $<\Delta Q^2(t)>$
of the accumulated heat fluctuation over a time span $t$ reads
\begin{equation}\label{variance}
<\Delta Q^2(t)> = < \Big( \int_0^t ds \Delta J^{\rm h} (s)\Big)^2 >
\end{equation}

Using a long measurement time span $t$ the time-dependent expectation value then relates to the zero frequency component of the PSD.
This is the case upon noting that the symmetrized correlation is a symmetric function of its argument and
assuming that the time-homogeneous auto-correlation of stationary heat fluctuations vanishes in sufficiently strong a manner
for infinite time. Then, the integral in ~(\ref{variance}) can be extended to infinity, yielding
\begin{equation} \label{Delta}
\lim_{t\rightarrow \infty} <\Delta Q^2(t)>/t  = \int_{-\infty}^\infty d\tau S^{\rm h}(\tau) = \tilde S^{\rm h}(\omega=0).
\end{equation}

The result for the zero-frequency limit therefore relates to a single time measurement of the manifest quantum observable $\Delta Q^2(t)$.
Still to measure accumulated `heat' rather than `heat-flux' presents a formidable challenge for the experimenter;
the case with accumulated electric charge is a lot easier accessible.
The detailed  behavior of this zero-frequency  nonequilibrium heat noise PSD will be studied next.

\begin{figure*}
\centerline{\includegraphics[width=0.75\textwidth]{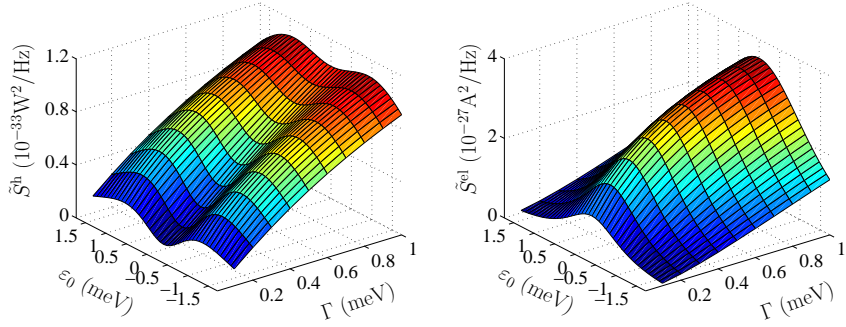}}
\caption{\label{fig:ga} (color online)
Power spectral density of the heat current noise at zero frequency $\omega=0$, (left panel) and
power spectral density of the electric current noise (right panel) as functions of the wire orbital site
energy $\varepsilon_0$ and wire-lead coupling strength $\Gamma$. The parameters employed are $T_{\rm L}= 6.2 K$ and $T_{\rm R}= 2.2K$.}
\end{figure*}

\subsection{Zero frequency noise power\\}

The theoretical PSD of heat current noise at zero frequency $\omega=0$ simplifies considerably, assuming the appealing form
\begin{align}\label{hnoise}
&\tilde{S}^{\rm h}(\omega=0; T_{\tl},T_{\tr})\notag\\
=&\frac{1}{2\pi\hbar}\int dE E^2 \{{\mathcal T}(E)( f_{\tl}(E)[1-f_{\tl}(E)]\notag\\
&+f_{\tr}(E)[1-f_{\tr}(E)])\notag\\
&+{\mathcal T}(E)[1-{\mathcal T}(E)][f_{\tl}(E)-f_{\tr}(E)]^2\}\;.
\end{align}
%This result is in agreement with a conjectured prediction in Ref.~\cite{krive01prb}.

Here the last line refers to a heat-shot-noise contribution while the first part corresponds to a nonequilibrium Nyquist-like  heat noise contribution.
Matters simplify considerably in thermal equilibrium where the shot noise contribution vanishes identically.

Let us also briefly contrast this result with the zero-frequency PSD of the fluctuations displayed by the nonlinear, accompanying thermoelectric current.
The latter reads~\cite{blanter00,kohler2005pr}:

\begin{align}
&\tilde{S}^{\rm el}(\omega=0; T_{\tl},T_{\tr})\notag\\
=&\frac{e^2}{2\pi\hbar}\int dE \{{\mathcal T}(E)(f_{\tl}(E)[1-f_{\tl}(E)]\notag\\
&+f_{\tr}(E)[1-f_{\tr}(E)])\notag\\
&+{\mathcal T}(E)[1-{\mathcal T}(E)][f_{\tl}(E)-f_{\tr}(E)]^2\}\;,
\end{align}

Most importantly, the zero-frequency PSD for heat current in Eq.~(\ref{hnoise}) differs by the energy factor $E^2$ within the integrand. Although this distinction seemingly appears minor and may even be guessed beforehand without going through the laborious task of doing a theoretical rigorous derivation from which this limit derives from the frequency-dependent main result given in Eq.~\eqref{longeqn}. It must be emphasized, however, that the two expressions lead to tangible differences. Particularly, note the different behavior
of the electronic and heat noise PSDs versus the tunable energy level $\varepsilon_0$ as depicted with Fig.~\ref{fig:curnoise} (b) and in Fig.~\ref{fig:ga}. While the zero-frequency component of the electric PSD at $\omega=0$ exhibits a maximum at $\varepsilon_0 = 0$, see in Fig. $2$(c) in Ref. \cite{Feiprb2011}, its heat current PSD possesses instead a local minimum at this value, see Fig.~\ref{fig:curnoise}(b). These two PSDs for charge current and heat current are compared in Fig.~\ref{fig:ga} over wide regimes of the electronic orbital energy $\varepsilon_0$ and the lead-molecule strength $\Gamma$.

These differences originate from the salient feature that the two transport mechanisms for charge and the energy are different.
The electric current is quantized by the electron charge ${e}$ while, in contrast, the energy carried by the electron is continuous and can assume principally an arbitrary value. Notably, the main contribution to the electronic noise power across the wire stems from those electrons occupying energy levels around the chemical potential $\mu = 0$. When $\varepsilon_0$ deviates from the chemical potential, increasingly less electrons participate in the transport. The flow of electron becomes diminished, and since both, the electric current and the electric noise are insensitive to the electron kinetic energies, they both decrease with increasing $|\varepsilon_0|$. This scenario differs for heat flow: There, the deviation from the chemical potential increases the possibility that successive electrons will carry different energies. This in turn causes an increase of
heat current noise. With further deviation of the orbital energy from the chemical potential, the occupancy difference [$f_{\rm L}(E) - f_{\rm R}(E)$] decreases monotonically; consequently the  noise power  $\tilde S^{\rm h}(\omega=0)$ decreases again.\\

\subsection{Electronic heat current noise in thermal equilibrium\\}
Next, let us focus on thermal equilibrium which is  attained when the two temperatures are
set equal, i.e. if $T_{\rm L}=T_{\rm R}$. In this case the average heat current vanishes identically, while its fluctuations
remain finite. The zero-frequency spectra of both noise spectra for heat and electric current noise increase upon
increasing the coupling strength $\Gamma$. This is so because the transmission probability increases.
The corresponding heat noise power is nonzero in equilibrium, however, as depicted with Fig.~\ref{fig:curnoise}(b).

In thermal equilibrium with $T_{\rm L}=T_{\rm R}=T$  the nonequilibrium zero frequency PSD  in Eq.~(\ref{hnoise}) simplifies further, obeying
\begin{align}
&\tilde{S}^{\rm h}(\omega=0, T) = 2 k_{\rm B} T^2 \tilde {G}^{\rm h}(\omega=0)\;,
\label{zero-FDT}
\end{align}
where $\tilde {G}^{\rm h}(\omega=0)$ denotes the static, linear heat conductance, obtained from expanding the result in Eq.~(\ref{current}) around a
small temperature bias and comparing with Eq.~(\ref{hnoise}).
This result is therefore in agreement with the fluctuation-dissipation-theorem (FDT) for the static heat conductance.

Note that an extension to a Green-Kubo-like, but now frequency dependent conductance, however, would intrinsically require also intermediate time-varying temperatures $T(t)$. Such a concept with a time-dependent, nonequilibrium temperature, cannot be justified  in the coherent quantum regime of an open system with only one level $\epsilon_0$  connecting the two leads. In fact even for  a different setup with a spatially   extended intermediate thermal conductor it has been found in Ref.~\cite{averin10prl} that at finite frequencies $\omega$ the PSD is not related to the corresponding  linear heat conductance in the ballistic, low temperature transport regime. This violation of the FDT is thus far from being fully settled in the literature.

The properties at zero absolute temperature, $T_{\rm L}=T_{\rm R}=0$, become even more subtle.
Here, the heat current PSD at finite frequencies $\omega$ still depends on frequency.
This dependence originates from quantum fluctuations where virtual transitions of
electrons  from lead-to-lead occur \cite{averin10prl}.
The Fermi distribution equals the Heaviside step function in this case. Therefore,
the contributions to the integrand in Eq.~(\ref{longeqn}) stems from the interval $[-\Omega,~0]$. After an integration of Eq. (\ref{longeqn}), one finds for the frequency dependent PSD the expression:
\begin{align}
&\tilde{S}^{\rm h}(\omega,T_{\tl}=T_{\tr}=0)\notag\\
=&\frac{\G}{4\pi\hbar}\left\{\left[(2\Omega)^2-2\G^2\right]\arctan\left(\frac{\Omega}{\G}\right)\right.\notag\\
&+2\left.\Omega\G\left[1+\log\left(\frac{\G^4}{\left(\Omega^2+\G^2\right)^2}\right)\right]\right\},~~\Omega \equiv \hbar\omega.\label{zero}
\end{align}

In the limit $\Gamma \rightarrow \infty$ the zero-temperature PSD scales like $\tilde{S}^{\rm h}(\omega) \propto \omega^3$.
This is in full agreement with  results obtained in  the work \cite{averin10prl} for a different setup, where
such an asymptotic behavior is found uniformly throughout the whole frequency region.
This uniform feature no longer holds true when $\Gamma$ is finite: The second term in the rhs of ~(\ref{zero}) introduces a linear cutoff in
the limit  $\omega \rightarrow 0$, so that $[\tilde{S}^{\rm h}(\omega)- \tilde{S}^{\rm h}(\omega=0)] \propto \omega$ in the extreme low frequency limit.
In distinct contrast, in the high-frequency region, the first term in the rhs of (\ref{zero}) becomes dominating.
As a consequence, the PSD~(\ref{zero}) in the high frequency limit approaches a square-law asymptotic crossover dependence, $\tilde{S}^{\rm h}(\omega) \propto \omega^2$.

\section{Conclusions and sundry topics\\}

By using the Green function formalism we have investigated electronic heat current. Our focus centered on the issue of the heat current fluctuations in a molecular junction model composed of a single orbital molecular wire. For the noninteracting case we succeeded in deriving a closed form for the frequency dependence of heat current noise; i.e. the
heat noise PSD, both in nonequilibrium $T_{\rm L} \neq T_{\rm R}$ and in thermal equilibrium $T_{\rm L}=T_{\rm R}$. The dependence of the heat current noise on the orbital energy $\varepsilon_0$ is qualitatively different from that for the accompanying electric current noise, see Fig.~\ref{fig:ga}. Moreover, the heat current fluctuation properties
depend strongly on the the overall tunneling coupling strengths $\Gamma_{\rm L}=\Gamma_{\rm R}= \Gamma$.

In the zero-temperature limit, the PSD of the heat current noise obeys two distinctive asymptotic behaviors,
being different in the intermediate-low frequency and in the high-frequency regimes. The particular square-law behavior of the PSD in the high-frequency region is due to the Lorentzian shape of the transmission
coefficient $\ma{T}(E)$ in Eq.~\eqref{transmission}.
Yet, the general effect would remain for any choice of the coefficient in the form of a localized, bell-shaped function: the noise spectrum will deviate from a cubic power-law asymptotic behavior upon entering the high-frequency region.

As emphasized in our introduction, with this work only the electron subsystem has been considered. Realistic heat transport in real molecular junctions would involve the complexity of interacting electrons and electron-phonon interactions \cite{dubi11rmp}. This electronic heat transport may dominate in certain situations so that the measured heat noise can be attributed approximately to the electronic component only. The unified approach, which would include both the electron and the phonon subsystems, as well as the effects of their interactions, presents a future challenge although several contributions in this direction for the average heat current (but not the heat current noise PSD) have already been undertaken before \cite{dubi11rmp,galperin07prb,galperin07jpcm,segal03jcp}.

\subsection{Open issues \\}
We conclude this study with further remarks that may shed light on challenging open problems and in addition may
invigorate others to pursue future research in  objectives addressed with our study. A first observation is that we obtained within the Green
function analysis tractable expressions for quantum transport in the steady state without ever having to invoke the
explicit knowledge of the inherent nonequilibrium density operator. Naturally, the quantum averages for the current
and the auto-correlation of the quantum fluctuations carry less information as encompassed with the full steady state
nonequilibrium density operator. The latter nonequilibrium density operator is typically very difficult to obtain and
explicit results are known for tailored situations only. In fact, explicit results are very intricate already for those
cases with overall quadratic Hamiltonians \cite{Dhar2012}.

As discussed above, a much more subtle issue refers to the experimental detection of quantum correlations.
In clear contrast to the case with a quantum, single-time expectation of a quantum observable, the issue of measurement
of manifest quantum correlations is a delicate  objective that is only rarely addressed with sufficient care in
the literature. This is so because the mere calculation of a theoretical two-time quantum correlations does not say anything
about its feasible experimental measurement scenario. Either strong (i.e. von Neumann-type) or weak quantum measurements impact
the dynamics as clearly manifested with the example of  the Zeno-effect \cite{Zeno1,Zeno2}.

With more than one time present
this objective relates to the problem of measurements of quantities that are not given in terms of quantum
observables \cite{Campisiprl,Campisipre,Talkner2007,clerk2010}. To appreciate the complexity somewhat in more detail let us
first consider the case with classical random variables. Then the PSD can be obtained
experimentally as the limit of a time average of the {\it classical} random process $J^{\rm h}(t)$, via considering the expression
\begin{equation}
\label{classicalS}
\tilde S^{\rm h}_{t\rightarrow \infty}(\omega) = \text{lim}_{t\rightarrow\infty} \frac{1}{2t} \left| \int_{-t}^{t} ds\, J^{\rm h} (s) \exp (i\omega s) \right|^{2} \;.
\end{equation}
Note that classically the measurement of the \textit{stochastic} variable of the instantaneous heat flow $J^{\rm h}(t)$  presents
no serious  problem while the same is not  straightforward for a quantum dynamics.
Moreover, even classically, the result in  (\ref{classicalS}) holds true only when the stochastic, finite value $\tilde S_{t}^{\rm h}(\omega)$ tends to the exact ensemble averaged value $\tilde S^{\rm h}(\omega)$, with its variance approaching zero as $t\rightarrow\infty$. The latter implies conditions of higher, fourth-order correlations to be satisfied \cite{Papoulis}. With the feature of dealing with the non-commutation property of
quantum observables at different times no such direct scenario is available for  experiment. Here the complexity of quantum
measurements will enter in its full generality. Only for tailored situations this task may simplify further, as it was the case in Sects. 3.2, 3.4 for the zero frequency  limit.

As mentioned already above, the case of quantum linear response theory may come as support also for nonequilibrium: The measurement
of a single observable (here the heat flux operator) due to an external perturbation is typically related to the evaluation of a specific
quantum correlation function \cite{HanggiPR82}. The case of the quantum-dissipation relation of Callen-Welton in thermal equilibrium
presents such a celebrated case \cite{callenwelton,cloizeaux,hanggi2005}. There, the dissipative part of the measurable,
frequency-dependent susceptibility of a perturbed observable $B$ is uniquely related to the power spectral density  $S_{BA}(\omega)$ of
quantum fluctuations of the observable $B$ and the fluctuations of observable $A$ to which an applied external conjugate force couples.
In our case it remains therefore a formidable task to identify the corresponding variable for the nonequilibrium situation so that
the single-time measurement of its linear response  becomes related to the heat PSD in Eq.~\eqref{longeqn} in a prescribed manner.
This at best is possible for the thermal equilibrium PSD in which an imposed energy perturbation couples to the thermal affinity $\Delta T/T$;  cf. in Refs. ~\cite{visscher,allen,liu}. This is not possible, however, for the equilibrium heat flow fluctuations at absolute $T=0$, with the inherent thermal affinity being divergent. In presence of quantum coherence destroying phenomena, such as high temperature or disorder, the nature of quantum correlations becomes suppressed. Then, the classical scenario can be used again to validate the theoretical predictions in thermal equilibrium \cite{blanter00,hanggi2005} and for tailored  steady-state nonequilibrium situations; note the nonequilibrium fluctuation theorems in Ref.~ \cite{HanggiPR82}.

%There is an intriguing perspective to apply an external periodic perturbation with the goal to control the spectral properties of the heat noise, %similar to
%electronic shot-noise control in ac-driven nanoscale conductors \cite{camalet04prb}. This idea can be realized, for example, by subjecting the %molecular wire to
%strong laser radiation \cite{kohler07naturenano} or by using direct modulations of the gate voltage. We conjecture that the role of laser radiation %may give
%rise to novel phenomena to be explored further by combining a Floquet theory for the driven system with the nonequilibrium Green function %formalism
%\cite{camalet04prb,kohler2005pr,rey07prb}.

\begin{acknowledgement}
Work supported by the German Excellence
Initiative via the ``Nanosystems Initiative Munich'' (NIM) (P.H.), the DFG priority
program DFG-1243 ``Quantum transport at the molecular scale'' (F.Z.,
P.H.), and the Volkswagen Foundation (Project No. I/83902) (P. H. and S.D).
\end{acknowledgement}

\end{document}